\documentclass[twocolumn,
 superscriptaddress,
 amsmath,amssymb,
 aps,
 floatfix
]{revtex4-2}

\usepackage{amsmath, amssymb, amsthm}
\usepackage[dvipsnames]{xcolor}
\usepackage{graphicx}
\usepackage{dcolumn}
\usepackage{bm}
\usepackage{mathrsfs}
\usepackage{hyperref}
\usepackage[capitalise]{cleveref}
\usepackage{tabularray}

\newcommand{\eexp}[1]{e^{#1}}
\newcommand{\norm}[1]{\left\lVert#1\right\rVert}

\begin{document}

\title{Dynamics of An Information Theoretic Analog of Two Masses on a Spring}

\author{Geoff Goehle}
\email{goehle@psu.edu}
\author{Christopher Griffin}%
\email{griffinch@psu.edu}
\affiliation{
	Applied Research Laboratory,
    The Pennsylvania State University,
    University Park, PA 16802
    }%

\date{\today}

\begin{abstract} In this short communication we investigate an information theoretic analogue of the classic two masses on spring system, arising from a physical interpretation of Friston's free energy principle in the theory of learning in a system of agents. Using methods from classical mechanics on manifolds, we define a kinetic energy term using the Fisher metric on distributions and a potential energy function defined in terms of stress on the agents' beliefs. The resulting Lagrangian (Hamiltonian) produces a variation of the classic DeGroot dynamics. In the two agent case, the potential function is defined using the Jeffrey's divergence and the resulting dynamics are characterized by a non-linear spring. These dynamics produce trajectories that resemble flows on tori but are shown numerically to produce chaos near the boundary of the space.  We then investigate persuasion as an information theoretic control problem where analysis indicates that manipulating peer pressure with a fixed target is a more stable approach to altering an agent's belief than providing a slowly changing belief state that approaches the target.   
\end{abstract}

\maketitle

\section{Introduction and Background}
The free energy principle was developed by Friston, Kilner and Harrison \cite{FKH06} as a thermodynamically inspired model of perception that presupposes the brain minimizes surprise (entropy) by maintaining an internal model of reality that is updated regularly through sensory input. Connections between this work, statistical mechanics and nonlinear dynamics have been subsequently studied by Friston and others \cite{F09,F10,KF11,LB13,KR15,BKMS17,BKR18,RBF18,CZCC21,AMTB22,FDSH23} and extended to more generic biological systems \cite{K12} as well as machine learning contexts \cite{MVCD22}. More recently, Hein et al. \cite{HMDM+24} showed that the principle of ``surprise minimization'' allows for collective behavior to arise in complex interacting systems.

In this short communication, we rephrase the free energy principle in terms of free \textit{entropy}, which arises naturally from the information geometry of the Fisher metric \cite{L61,PV01,B03}. This quantity and the corresponding techniques from information geometry have been used extensively in statistical physics \cite{B11,SSBC12,SC12,KLHH16,FC09,KH17,K21,GPB20,NGCG20}. 

Using techniques from classical mechanics on manifolds, we derive an alternative to the DeGroot averaging model (see e.g., \cite{DGL13,SGSR18,GSJ22}), which has similar structure and properties but emerges naturally from our information theoretic variation of the free energy principle as a classical mass/spring system, further supporting the hypothesis that ``surprise minimization'' can lead to collective behavior \cite{HMDM+24}. In this derivation, a statistical divergence replaces the classical square distance in the Hooke spring potential. Convergence to an opinion is ensured by the inclusion of a dissipative ``dashpot'' term \cite{M15,BC21} and consensus is determined by the spring constants. While opinion dynamics have been studied extensively \cite{DeGroot74,Krause00,HK02,BN05,WDA05,Toscani06,Weisb06,Lorenz07,BHT09,CFL09,KR11,DMPW12,CFT12,JM14,SZAS21,GG21} along with the related disciplines of flocking and consensus \cite{DeGroot74,Krause00,Centola15,TT98,CS07,EK01,L08,LX10, DM11,MT14,GSJ22}, we believe this is the first derivation of these dynamics from the free energy principle or the corresponding ``minimum surprise principle'' rephrased as a free entropy minimization. 

Let $p(\mathbf{x}|\bm{\eta})$ be a probability distribution parameterized by $\bm{\eta}$. For two distributions $p(\mathbf{x}\vert\bm{\eta})$ and $q(\mathbf{x}\vert\bm{\eta})$, the Kullback-Liebler (KL) divergence is given by,
\begin{equation}
D_{KL}(p\vert q) = \int_X p(\mathbf{x}\vert\bm{\eta}) \log\left(\frac{p(\mathbf{x}\vert\bm{\eta})}{q(\mathbf{x}\vert\bm{\eta})}\right) d\mathbf{x}.
\end{equation}
It is straightforward to see that the KL-divergence is neither symmetric nor does it satisfy any version of the triangle inequality \cite{nielsen2018}. 
Likewise, the quadratic form of the Fisher metric is defined componentwise as,
\begin{equation*}
g_{jk}(\bm{\eta}) = \int_X \frac{\partial \log\left[p(\mathbf{x}|\bm{\eta})\right]}{\partial x_j}\frac{\partial \log\left[p(\mathbf{x}|\bm{\eta})\right]}{\partial x_k}p(\mathbf{x}|\bm{\eta})\,dx.
\end{equation*}
In addition to the usual Riemannian geodesic distance, statistical manifolds admit divergences, which are locally consistent with the Fisher metric and yield affine connections \cite{A10, nielsen2018}, but are themselves not true distances. The KL-divergence, already introduced, is one such divergence. In particular, we can locally write,
\begin{equation*}
D_{KL}\left(p(\mathbf{x}\vert\bm{\eta})\vert p(\mathbf{x}\vert\bm{\eta}_0)\right) =
\sum_{j,k}\frac{1}{2} \mathbf{g}_{jk}(\bm{\eta}_0)\Delta\eta^j\Delta\eta^k + O(\norm{\bm{\Delta\eta}}^3),
\label{eqn:fisherkl}
\end{equation*}
showing the relationship between the metric and the divergence \cite{nielsen2018}.  In this communication we use the Fisher metric to measure distance and length locally, while the KL-divergence will play the role of a global Mahalanobis distance.

Using the kinetic energy, computed using the Fisher metric, we can form a Lagrangian
\begin{equation}
    L(\bm{\eta},\dot{\bm{\eta}}) = \sum_{j,k} \frac{1}{2}\mathbf{g}_{jk}(\bm{\eta}) \dot{\eta}^j \dot{\eta}^k - V(\bm{\eta)},
    \label{eqn:GeneralLagrangian}
\end{equation}
where $V$ is a potential function, to be defined in context.

Let $\Delta_{n-1}$ be the $n-1$ dimensional unit simplex embedded in $\mathbb{R}^n$. The categorical distribution with parameters $q^1,\dots,q^{n-1}$ is the vector $(q^0,\dots,q^{n-1}) \in \Delta_{n-1}$, where
\begin{equation}
    q^0 = 1 - \sum_{j} q^j.
\end{equation}
In this case, the metric tensor can be computed explicitly as \(g_{ij} = \delta_{ij}/{q^i} + 1/q^0,\) where $1 \leq i,j\leq n-1$.

\section{Opinion Dynamics as a Nonlinear Spring System}
Consider the two outcome (Bernoulli) case and let $q^1 = q$ and $q^0 = 1-q$. Then the kinetic energy is,
\begin{equation}
T(q) = \frac{m}{2}\frac{\dot{q}^2}{q(1-q)}.
\label{eqn:KineticEnergy}
\end{equation}
Here, $m$ is a mass term that could be unitized.

For the remainder of this paper, we consider the setting where $q$ is the parameter in a Bernoulli random variable representing an agent's belief about a certain outcome or opinion about a certain bivalent topic. If $q' \in (0,1)$ is a fixed belief point (i.e., an external reference point) then the KL-divergence yields an information theoretic ``squared distance'' from $q$ to $q'$ as,
\begin{equation*}
    D(q,q') = q\log\left(\frac{q}{q'}\right) + (1-q)\log\left(\frac{1-q}{1-q'}\right).
\end{equation*}
In this scenario $q'$ is treated as a prior belief and $q$ is treated as a posterior belief. Mimicking the potential energy for a spring, and using \cref{eqn:GeneralLagrangian} and \cref{eqn:KineticEnergy}, the Lagrangian written in terms of $q$, $\dot{q}$, and $q'$ is,
\begin{equation*}
L(q,q') = \frac{m}{2}\frac{\dot{q}^2}{q(1-q)} - \frac{k}{2} D(q,q'),
\end{equation*}
where we assume (for now) that $q'$ represents a fixed Bernoulli distribution and $k$ is a notional spring constant. This has corresponding Hamiltonian,
\begin{equation*}
    H(q,p) = \frac{q(1-q)}{2m}p^2 + \frac{k}{2} D(q,q'),
\end{equation*}
where,
\begin{equation*}
    \frac{\partial L}{\partial \dot{q}} = p = \frac{m\dot{q}}{q(1-q)},
\end{equation*}
is the conjugate momentum variable. Expanding to second order around $q = q' = \tfrac{1}{2}$, $H(q,p)$ has form,
\begin{equation*}
    H(q,p) = \frac{q(1-q)}{2m}p^2 + k (q-q')^2 + O(q^3),
    \label{eqn:HamiltonianQ}
\end{equation*}
showing that this is just a nonlinear spring potential. 


Using the foregoing analysis, we can derive a variant of the DeGroot dynamics \cite{SGSR18,G21,GSJ22}. As before, let $q_i$ be the parameter in a Bernoulli random variable representing an agent's belief about a certain outcome or opinion about a certain bivalent topic. Let $k_{ij}$, be a spring constant, denoting the degree to which agent $i$ and agent $j$ influence each other's belief and let $q_i'$ be a prior belief held by agent $i$ with a restoring force $k_i$. Then the Hamiltonian for this system in canonical coordinates is given by,
\begin{multline}
H(\mathbf{q},\mathbf{p}) = \sum_{i}\frac{q_i(1-q_i)}{2m_i}p_i^2 + \sum_{i}\frac{k_i}{2}\left(q_i - q'_i\right)^2 +\\ \sum_{i<j} \frac{k_{ij}}{2}\left(q_i - q_j\right)^2 + O\left(\norm{\bm{q}}^3\right).
\label{eqn:NewModel}
\end{multline}

Assuming each agent experiences stress from an opinion that differs from a fixed (or prior) opinion $q'_i$, the standard DeGroot Hamiltonian (stress function) \cite{SGSR18,G21,GSJ22} is similar,
\begin{equation}
H_D(\mathbf{q}) = \sum_{i}\frac{k_i}{2}\left(q_i - q'_i\right)^2 + 
\sum_{i<j} \frac{k_{ij}}{2}\left(q_i - q_j\right)^2.
\label{eqn:DeGroot}
\end{equation}
Under most circumstances, $q_i'$ is the initial belief of the agent, although it can represent any fixed belief that influences the agent. The repeated averaging dynamics \cite{SGSR18,GSJ22} are derived by differentiating $H_D$ with respect to $q_i$ and solving for the local minimum. Agents then update their belief to the local minimum. Alternatively, Gavrilets et al. \cite{GAV16} assumes a Jacobi iteration so that, 
\(    \dot{q}_i = -\frac{\partial H_D}{\partial q_i},\)
which is consistent with the consensus dynamics literature (see \cite{MT14} for an overview). 

A modification to the stress function that is consistent with the free energy hypothesis \cite{FKH06} is to assume that the stress of internal model adjustment occurs continuously, rather than from a fixed initial assumption. This adds an additional term to \cref{eqn:DeGroot}, 
\begin{multline}
\hat{H}_D(\mathbf{q}) = \sum_{i}\frac{\hat{k}_i(\epsilon)}{2}\left[q_i(t+\epsilon) - q_i(t)\right]^2 \\ + \sum_{i}\frac{k_i}{2}\left(q_i - q'_i\right)^2 +
\sum_{i<j} \frac{k_{ij}}{2}\left(q_i - q_j\right)^2, 
\label{eqn:DeGrootRate}
\end{multline}
and allows the stress contributed by the initial belief to be disregarded (by setting \(k_i=0\)) while still producing transitive dynamics.  If $\hat{k}_i(\epsilon) \propto 1/\epsilon$ then we recover the form of \cref{eqn:NewModel} from \cref{eqn:DeGrootRate} as $\epsilon \to 0$.

\begin{figure*}[htbp]
\centering
\includegraphics[width=0.31\textwidth]{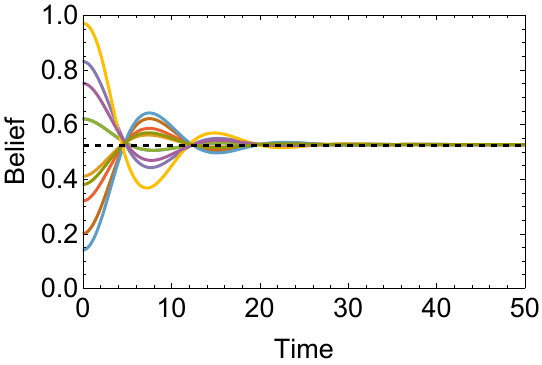}\quad
\includegraphics[width=0.31\textwidth]{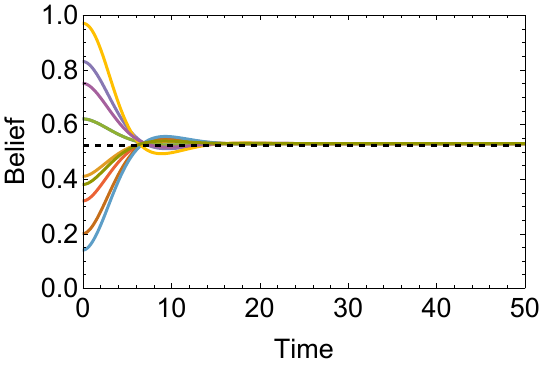}\quad
\includegraphics[width=0.31\textwidth]{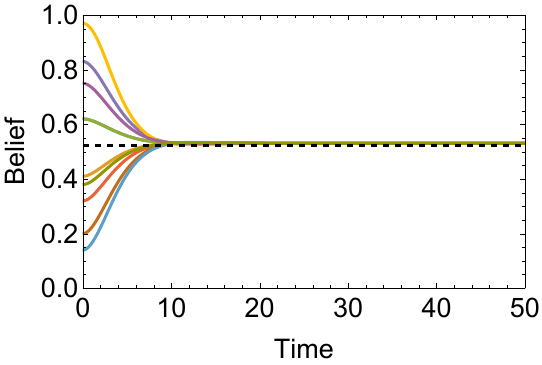}
\caption{When $k_i = 0$ for all $i$, agents converge to consensus as a result of increasing peer-pressure in the dissipative system. Prior to convergence, the dynamics oscillate consistent with Fink et al.'s theory of oscillating belief. (Left) $\alpha = 0.35$. (Center) $\alpha=0.6$. (Right) $\alpha=0.75$. The mean of the initial conditions is shown as a dashed line.}
\label{fig:TenAgents}
\end{figure*}

The dynamics of \cref{eqn:NewModel} and \cref{eqn:DeGroot} are fundamentally different. The kinetic energy term causes belief oscillation (discussed in detail in the sequel), while this is not observed in long-run human behavior (though may be expected in the short term \cite{FKHM02}), with convergence to a consensus opinion expected over time. We can recover such dynamics by introducing a dissipative element to the Lagrangian used to generate \cref{eqn:NewModel}. As in \cite{BC21}, let,
\begin{equation*}
    \tilde{L}(q,q') = \eexp{-\alpha t}\left[ \frac{m}{2}\frac{\dot{q}^2}{q(1-q)} - \frac{k}{2} D(q,q')\right].
\end{equation*}
Then the Hamiltonian of the system becomes,
\begin{equation*}
    \tilde{H}(q,q') =\eexp{-\alpha t}\frac{q(1-q)}{2m}p_i^2 + \eexp{\alpha t}\frac{k}{2} D(q,q').
\end{equation*}
Passing to the general case, our proposed model is,
\begin{multline}
\tilde{H}(\mathbf{q},\mathbf{p}) = \eexp{-\alpha t}\sum_{i}\frac{q_i(1-q_i)}{2m_i}p_i^2 +\\ \eexp{\alpha t }\left[\sum_{i}\frac{k_i}{2}D(q_i,q_i') +\sum_{i\neq j} \frac{k_{ij}}{2}D(q_i,q_j)\right].
\label{eqn:NewModelA}
\end{multline}
Like the repeated averaging DeGroot model with increasing peer-pressure \cite{SGSR18,G21,GSJ22}, this system converges to consensus when $k_i = 0$ for all $i$; that is, when there are no external spring forces outside of the agents' beliefs. The oscillating behavior for smaller values of $\alpha$ is  consistent with some theories of human cognition  (see Fink et al. \cite{KFWV96, FKHM02} and their references), which assert that beliefs will oscillate, going so far as to use a spring analogy.  Convergence to consensus is illustrated in \cref{fig:TenAgents} for varying values of $\alpha$. It is worth noting that when $\alpha$ is small, convergence does not necessarily occur to a weighted mean of the initial conditions, as it does in the work of Gavrilets et al. \cite{GAV16} and in Griffin et al. \cite{SGSR18,G21,GSJ22}.

\begin{figure}[htbp]
\centering
\includegraphics[width=0.65\columnwidth]{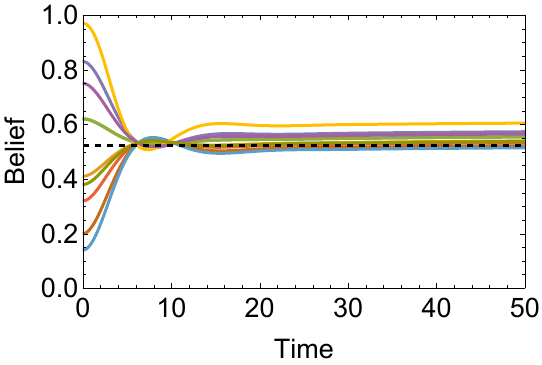}
\caption{When $k_i > 0$ for some $i$, consensus is not ensured and agents will converge to non-consensus states. Here $\alpha=0.5$ and $k_i = 0.3$ for all $i$. The mean of the initial conditions is shown as a dashed line.}
\label{fig:Converge}
\end{figure}

Notably when $k_i > 0$ for some $i$, the inclusion of fixed external belief forces implies that consensus is not ensured, even though the agents converge to fixed beliefs, as is expected for a dissipative system. This is shown in \cref{fig:Converge}, where $k_i = 0.3$ for all $i$.

For practical modeling purposes, we expect the mass terms $m_i$ and $\alpha$ to be comparatively large, explaining why decision makers are slow to make changes to their mental models and do not oscillate substantially in their beliefs.  This reduces the kinetic energy component of the Hamiltonian, thus making the model much more like the DeGroot model with an increasing peer-pressure term as in \cite{GSJ22}. Notably the dissipative multiplier is equivalent to assuming jointly increasing peer-pressure and decreasing mass. Nevertheless, it is worth studying the Hamiltonian dynamics in non-dissipative systems as an interesting example of a nonlinear spring-mass system in an unusual geometry. This investigation can also provide insight into transient belief dynamics.

\section{Chaotic Dynamics in the Two Agent System}
Consider the two-agent Hamiltonian,
\begin{equation}
    H(\mathbf{q},\mathbf{p}) = \frac{1}{2}q_1(1-q_1)p_1^2 + 
    \frac{1}{2}q_2(1-q_2)p_2^2 + \\ \frac{k}{2}D_J(q_1,q_2),
\label{eqn:TwoAgents}
\end{equation}
where mass constants are unitized for simplicity and $D_J(q_1,q_2)$ is the Jeffrey's divergence,
\begin{equation*}
    D_J(q_1,q_2) = D(q_1,q_2) + D(q_2,q_1).
\end{equation*}
This is a classic two-mass spring system in a non-Euclidean geometry. In the Euclidean case, it is easy to see the resulting dynamics are integrable and admit two constants of motion, the energy (given by the corresponding Hamiltonian) and the linear momentum of the total system. In contrast, we can show numerically that the Hamiltonian dynamics in \cref{eqn:TwoAgents} are not integrable, showing signs of weak chaos \cite{Z92}. 

\begin{figure}[htbp]
\centering
\includegraphics[width=0.65\columnwidth]{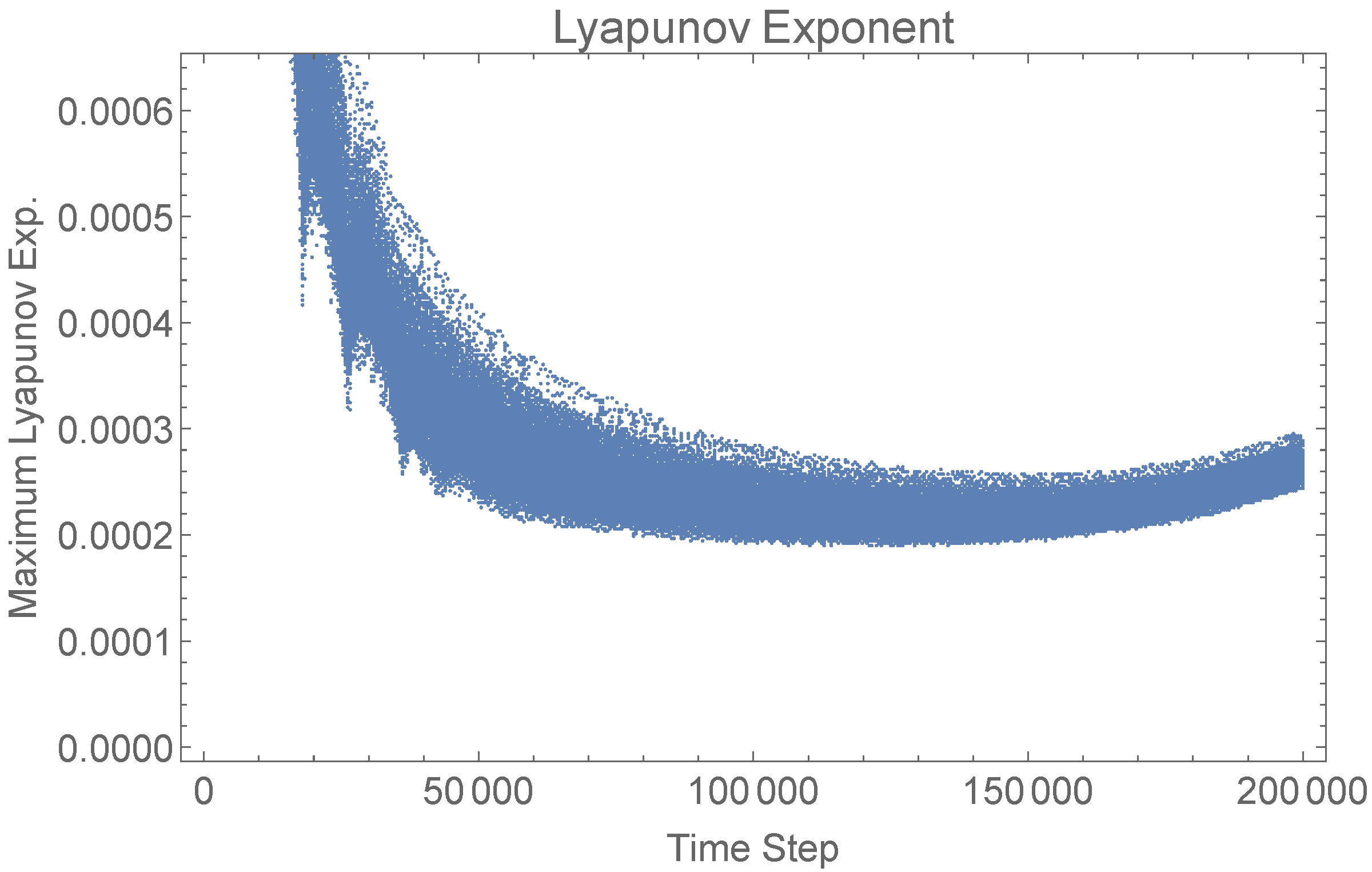}
\caption{The Wolf algorithm \cite{WSSV85} is used to estimate the Lyapunov spectrum for the Hamiltonian dynamics arising from \cref{eqn:TwoAgents}. The upturn in the plot of the estimate for the maximum Lyapunov exponent suggests weak chaos caused by behavior near the boundaries of the space. The Wolf algorithm estimate for the maximum Lyapunov is $\lambda \approx 0.0003 > 0$.}
\label{fig:Lyapunov}
\end{figure}

Introduce the canonical coordinate transform,
\begin{equation}
    q_i = \cos^2\left(\frac{\theta_i}{2}\right) \qquad i\in\{1,2\}.
    \label{eqn:Diffeomorphism}
\end{equation}
Then \cref{eqn:TwoAgents} becomes a classical non-linear spring Hamiltonian with two coupled masses,
\begin{equation}
    H(\bm{\theta},\bm{\omega}) = \frac{p_1^2}{2} + \frac{p_2^2}{2} + \frac{k}{4}\left[\cos(\theta_1) -  \cos(\theta_2)\right]\log\left[\frac{\tan^2\left(\frac{\theta_2}{2}\right)}{\tan^2\left(\frac{\theta_1}{2}\right)} \right].
    \label{eqn:twoSpringTheta}
\end{equation}
As expected and consistent with \cref{eqn:NewModel}, the second order approximation for this is,
\begin{equation}
H_0(\bm{\theta},\bm{\omega}) = \frac{p_1^2}{2} + \frac{p_2^2}{2} + \frac{k}{2}(\theta_1 - \theta_2)^2 + O(\norm{\bm{\theta}}^3).
\label{eqn:twoSpringThetaApprox}
\end{equation}
Consequently, we can express $H(\bm{\theta},\bm{\omega})$ as,
\begin{equation*}
   H(\bm{\theta},\bm{\omega}) = H_0(\bm{\theta},\bm{\omega}) + \epsilon H_1(\bm{\theta},\bm{\omega}),
\end{equation*}
where $H_0$ is an integrable Hamiltonian, $\epsilon$ is a perturbation constant and $H_1$ is the nonlinear Hamiltonian perturbation. Since the coordinate transform in \cref{eqn:Diffeomorphism} is a diffeomorphism, we expect to see Hamiltonian chaos arise in the dynamics generated by \cref{eqn:TwoAgents}. We find evidence for weak chaos \cite{Z92} associated to behavior near the manifold boundary. To obtain this result, we used a time-varying estimate for the maximum Lyapunov exponent of the dynamical system using the Wolf algorithm \cite{WSSV85}. In particular, we used Sandri's implementation of the Wolf algorithm \cite{S96}, which returns an estimation of the the Lyapunov spectrum using a single long trajectory after an initial burn-in period. \cref{fig:Lyapunov} shows the instantaneous estimates of the maximum Lyapunov exponent during algorithm execution over the long trajectory. The final estimate provided by the algorithm for the maximum Lyapunov exponent is $\lambda \approx 0.0003 > 0$. Positivity of the maximum Lyapunov exponent is further supported by the fact that the trajectory shown in \cref{fig:Lyapunov} turns up after an initial period of decay. In \cref{fig:Lyapunov}, the spring constant $k = 2$, which improves numerical stability of the long trajectory required to execute the Wolf algorithm. The fact that the maximum Lyapunov exponent is positive and the dynamics play out on a compact set is sufficient to conclude that they are (weakly) chaotic.

\begin{figure}[htbp]
\centering
\includegraphics[width=0.65\columnwidth]{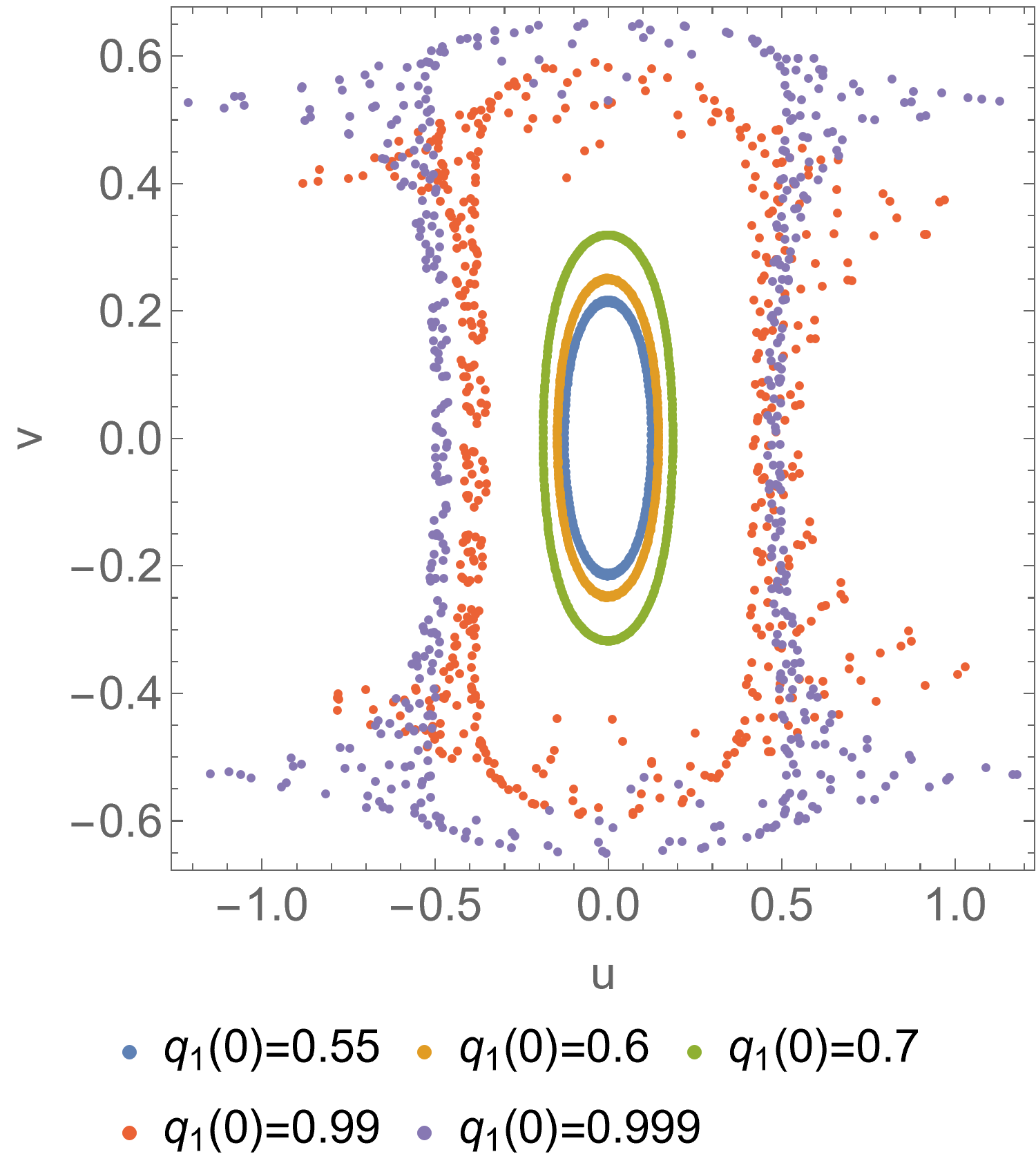}
\caption{Projection of solutions to \cref{eqn:twoSpringTheta} to the plane spanned by \(u =\left \langle{0,0,1} \right \rangle\) and \(v =  \langle  \frac{1}{\sqrt{2}},-\frac{1}{\sqrt{2}},0 \rangle\). We fix $q_2(0)=\tfrac{1}{4}$ and vary $q_1(0)$. As the initial condition for $q_1$ is varied, the structure of the trajectory $(q_1,q_2)$ changes. When $q_1(0)$ is close to $\tfrac{1}{2}$, the dynamics are in a spatial region that is largely flat and the trajectories are projected ellipses. When $q_1(0)$ is far from $\tfrac{1}{2}$, the geometry influences the trajectory and the trajectories are more chaotic.}
\label{fig:Spirals}
\end{figure}

This is consistent with prior results for Hamiltonian chaos on a compact set defined by the unit simplex that showed that the chaotic dynamics arose as a result of interactions with the boundary \cite{GSB22}. This Hamiltonian system is unusual in that it has an infinite line of fixed points $q_1 = q_2$ and $p_1 = p_2 = 0$. The result is a degeneracy, which makes application of the classical Kolmogorov–Arnold–Moser (KAM) theorem impossible \cite{A13} because the dynamics generated by $H_0$ evolve on a cylinder homeomorphic to $\mathbb{S}^1 \times \mathbb{R}$, rather than the two-torus $\mathbb{S}^2$. From this we see that sensitivity to initial conditions and the classical behaviors associated to weak chaos emerge as a result of asymmetries in the initial conditions. This is shown in \cref{fig:Spirals}, where we fix $q_2(0)=\tfrac{1}{4}$ and vary $q_1(0)$, illustrating the distinct trajectory structures that emerge. In \cref{fig:Spirals}, we project all trajectories onto the plane $q_1 + q_2 = 1$, effectively collapsing the cylinders on which the dynamics evolve near the interior fixed point $q_1 = q_2 = \tfrac{1}{2}$ into ellipses.
These nested cylinders are (essentially) degenerate KAM tori. This is because of the region in the interior of unit simplex, where the geometry is locally Euclidean. In this region, the behavior of the system is given almost entirely by a two-mass and spring system. Non-integrability (and thus the destruction of additional conserved quantities) is only introduced when the dynamics interact with the boundary, which induces an effective nonlinear force, not present in Euclidean space. That is, when $q_1(0)$ is near the boundary, we see a more chaotic trajectory emerge in \cref{fig:Spirals} a result of the boundary interactions.

\section{Control of Opinion Dynamics}
The introduction of nonlinear forces from the geometry of the space itself has implications for the corresponding control problem, which can provide insights into understanding persuasion in bivalent topics \cite{GG24}.  Of particular interest is the question of whether an external belief (i.e., perceived evidence) or its associated spring constant (i.e., social pressure) are more effective control parameters.  

Formally stated, given a binary belief in \(\theta\)-coordinates and an initial belief \(\theta_0\), the goal is to use either an external belief \(\theta'\) or the associated spring constant \(k\) to move \(\theta\) to a desired target \(\theta_f\) in time horizon \(T\) while minimizing the objective function, 
\begin{equation}
    \label{eqn:objective}
    \mathcal{O} = \int_{t_0}^{t_f}  \frac{p^2}{2} + C(k, \theta, \theta')\, dt,
\end{equation}
where \(C(k, \theta, \theta')\) is a cost associated to the control parameter.  The exact definition of \(C\) depends on the method of control (external belief or spring constant) and is specified in the following. We include the kinetic energy term in \cref{eqn:objective} because without some cost associated to momentum, \cref{eqn:objective} degenerates into a ``bang-bang'' style control problem \cite{JG63}.  Additionally, the momentum cost is related to the ``accumulated surprise'' generated by the control signal \cite{GG24}, and minimizing the accumulated surprise is consistent with the free energy principle. 

We work in $\theta$ coordinates for simplicity of form and because the resulting dynamics are more numerically stable. The associated control Hamiltonian (which is distinct from the system Hamiltonian \(H\)) \cite{K04} is,
\begin{equation}
\label{eqn:controlHamiltonian}
\mathcal{H} = \frac{p^2}{2} + C(k, \theta, \theta') + \lambda \frac{\partial H}{\partial p} - \mu \frac{\partial H}{\partial \theta},
\end{equation}
where \(\lambda\) and \(\mu\) are the Lagrange multipliers associated to \(\theta\) and \(p\), respectively.

For the \(\theta'\)-control case we define the cost function,
\[
C_D(\theta, \theta') = \beta D[\cos^2(\theta/2), \cos^2(\theta'/2)],
\]
and suppose that the state dynamics are derived from the full Hamiltonian \(H\) from \cref{eqn:twoSpringTheta} with \(\theta_1=\theta\) and \(\theta_2 = \theta'\).   The optimal control path resulting from \cref{eqn:controlHamiltonian} is characterized by Pontryagin's maximum principle \cite{S22}, as usual.  Since the resulting system of differential equations is poorly behaved, we also consider the second-order approximation to \(\mathcal{H}\) with cost function, 
\[
C_S(\theta, \theta')= \frac{\beta}{4}(\theta - \theta')^2,
\]
and state dynamics given by \cref{eqn:twoSpringThetaApprox}.  The approximate control problem is quadratic in cost with linear dynamics (LQC) and, consequently, is much better behaved \cite{K04}.  

As long as both belief boundary conditions are sufficiently removed from the edges of the space, the \(\theta'\)-control problem is well behaved and the second-order approximation is reasonably accurate. The system becomes poorly behaved numerically in regions near certainty because the curvature of the information manifold makes both the cost and the state dynamics stiff.  Additionally, the approximate control problem begins to exhibit wrapping in this regime, resulting in non-physical solutions. Since the control variable acts via the KL-divergence, which is unbounded on the manifold, it would seem possible to reach any target belief within an arbitrary time horizon.  However the Fisher-Rao metric is poorly conditioned near the manifold boundaries so that in practice the control problem becomes intractable, in addition to a potential breakdown in overall model fidelity. 

\begin{figure}[htbp]
\centering
\includegraphics[width=0.75\columnwidth]{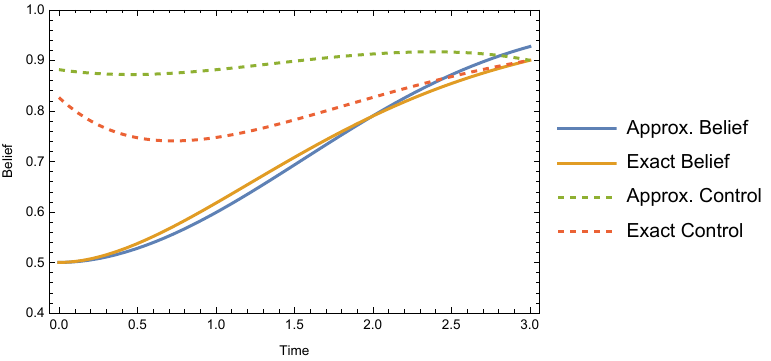}
\caption{Belief evolution and control for the \(\theta'\)-control formulation of \cref{eqn:objective}.  The plot shows solution curves for the exact formulation of the control Hamiltonian as well as the second-order approximation in belief coordinates \(q\).  The boundary values correspond to \(q_0 = 0.5\), \(p_0 = 0\) and \(q_f = 0.9\) with \(\alpha = 0.5, \beta = 10, k=1\) and time horizon \(T=3\). }  
\label{fig:UControl}
\end{figure}

To demonstrate, consider boundary values \(\theta(0)=2\arccos\sqrt{0.5}\) and \(\theta(3)=2\arccos\sqrt{0.9}\) (\(q_0=0.5\) and \(q_f = 0.9\), respectively) and parameters \(\alpha = 0.5\), \(\beta = 10\), and \(k=1\). Additionally assume zero initial momentum, \(p(0)=0,\) and the transversality boundary value \(\mu(3)=0\).  For the exact formulation of the control problem, a globally optimal solution can be found numerically using standard solvers.  For the approximate formulation, an approximately optimal control path can be written analytically, as it is a LQC problem \cite{K04}.  We can model the behavior of this solution under the full system dynamics by using the approximate control to propagate the forward initial value problem defined via \(H\). \cref{fig:UControl} shows the resulting control paths \(\theta'\) and corresponding belief evolution in probability coordinates \(q\) for both the exact and approximate control problems.  The two solutions are fairly similar with the approximate solution achieving a final belief value of \(q_f = 0.93\) instead of the intended \(q_f = 0.9\). 

Since the spring potential is smaller than the KL-divergence, particularly near the boundaries where the KL-divergence goes to infinity, we find that the approximate solution generally achieves a boundary value more extreme (closer to the boundaries of the manifold) than the desired target when applied to the exact dynamics.  This is an acceptable error in most applications, although it does come with the increased costs associated to using a non-optimal solution. 

While proximity to the geometric boundary drives stiffness in the control problem, this can be ameliorated to some extent by choosing a longer time horizon \(T\). We can very roughly approximate a lower bound for an acceptable time horizon by solving the approximate control problem with the maximum (minimum) possible control input \(\theta'=\pi\) (\(\theta'=0\)), to estimate the shortest horizon over which the desired belief target can be met without ``wrapping'' the control.  While this lower bound doesn't have a clean analytic formula, it is straightforward to compute numerically.  Using the parameter values above the minimum time horizon to avoid wrapping is \(T\approx 1.9\).  Numerical experimentation indicates that reasonable control solutions can be found for \(T > 2.2\), below which both the exact system and the forward simulation of the approximate control system become stiff.  As \(T\) approaches this threshold the approximate control curve gets closer to the boundaries of the geometry, and the approximation breaks down.   

As an alternative to controlling the belief distribution by changing \(\theta'\), we can instead fix \(\theta'=\theta_f\) to be the desired belief and use the spring constant \(k\) as the control parameter.  We use the same objective function as~\cref{eqn:objective} with the quadratic cost \(C_K(k) = \beta k^2\).  We use this cost for both the exact dynamics, given by \(H\), and approximate dynamics, given by \(H_0\).  The resulting \(k\)-control problems do not appear to have a simple closed-form solution.  However, the approximate control problem is again a LQC problem and both formulations are much better behaved numerically than the \(\theta'\)-control formulation.  This is because the control parameter \(k\) does not lie on the information manifold and is not as strongly impacted by the manifold curvature.  

\begin{figure}[htbp]
\centering
\includegraphics[width=0.65\columnwidth]{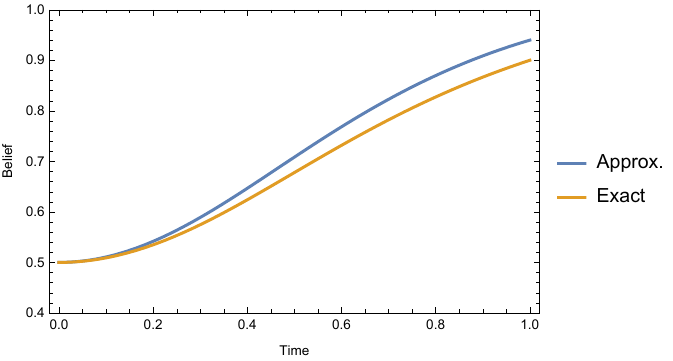} \\
\includegraphics[width=0.65\columnwidth]{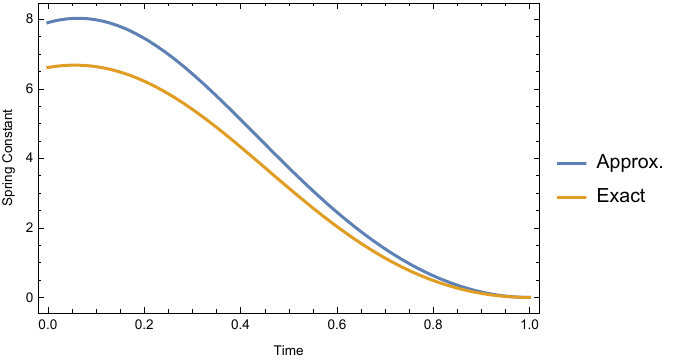}
\caption{Belief evolution (top) and control (bottom) for the \(k\)-control formulation of \cref{eqn:objective}.  The plots show solution curves for the exact formulation of the control Hamiltonian as well as the second-order approximation in belief coordinates \(q\).  The boundary values correspond to \(q_0 = 0.5\), \(p_0 = 0\) and \(q_f = 0.9\) with \(\alpha = 0.5, \beta = 10, \theta'=2\arccos\sqrt{0.9}\) and time horizon \(T=1\). }  
\label{fig:KControl}
\end{figure}

As an example, consider the same control problem as above except with \(\theta'=3\arccos\sqrt{0.9}\) and \(T=1\).  As before, we compute both the solution to the full control Hamiltonian as well as the approximate control Hamiltonian.  For the approximate solution we again recompute the forward problem using the exact dynamics.  The results are shown in~\cref{fig:KControl}. Again the second-order approximation provides reasonable control of the full system, although the approximate solution underestimates the response of the belief to the control variable for the same reasons as the \(\theta'\)-control case. The approximate solution produces a final belief of \(q_f=0.94\), instead of the intended target \(q_f = 0.9\). 

Notably the time horizon for this formulation of the problem is smaller than the minimum threshold estimated for the \(\theta'\)-control formulation.  Since \(k\) is not bounded by the information geometry there is no corresponding threshold in the \(k\)-control case and the numerical problem remains tractable down to \(T\approx0.1\). 

While a more thorough analysis of this information control problem, including joint \((k,\theta')\)-control, is outside the scope of this paper, this initial analysis indicates that the spring constant (i.e., social pressure) is a more robust control parameter than the observed belief.  At a high level this is because the spring constant produces information potentials which scale linearly, while the observed belief can only achieve large potentials by interacting with the highly curved portions of the manifold boundary.  Additionally, the simpler second-order approximation to the control problem, which utilizes traditional spring potentials and is a LQC problem, can produce effective control solutions, particularly in regions of high entropy.

\section{Conclusion and Future Directions}
In this paper, we introduced a model of belief combining Friston's minimum energy principle and information geometry, making use of the Bayesian brain hypothesis. Using a Bernoulli distribution as a model of an agent's belief about a bivalent topic (e.g., humans cause climate change), we showed how to recover a variant of DeGroot opinion dynamics via a classical Lagrangian. We further showed conditions under which opinion converges or comes to consensus through the use of a dissipative Lagrangian. In studying the nonlinear dynamics for two agents arising from the non-dissipative form of the Lagrangian, we showed that the corresponding Hamiltonian was a perturbation of a well-known integrable Hamiltonian. Despite the fact that the KAM theorem does not apply in this case, we illustrated symplectic flow near the fixed points and used Wolf's algorithm to numerically prove the existence of chaotic behavior in trajectories near the boundary. 

In studying the control problem arising from these dynamics, we identified interesting facts about the problem of persuasion. We found that manipulating peer pressure with a fixed (preferred) belief state is a more stable approach to altering an agent's belief than providing a slowly changing belief state that approaches a target. These results may have real-world implications. Prior research shows that peer-pressure is a powerful motivator for inducing cooperation \cite{MRP13}, especially (for example) in climate change beliefs and actions \cite{DD19,SPB19}. Moreover, while information innoculation appears to be the most effective counter-misinformation strategy, research supports the hypothesis that ``nudging'' (through social pressure) provides a potentially useful way to counter-act misinformation as opposed to ``debunking'', which has been shown to be less effective \cite{RCS23}. The mathematical results presented in this paper provide the first potential explanation for these empirical observations using a biophysical model. As such, further research into model generalizations may provide additional insights into persuasion and misinformation management. 

\section*{Acknowledgement}
This research was developed with funding from the Defense Advanced Research Projects Agency (DARPA) through NavSea Task Description DO 21F8366 and Contract HR0011-22-C-0038. The views, opinions and/or findings expressed are those of the author and should not be interpreted as representing the official views or policies of the Department of Defense or the U.S. Government. Distribution Statement ``A'' (Approved for Public Release, Distribution Unlimited).

\bibliography{../InformationOscillator/Oscillator}

\end{document}